\documentclass[9pt,twocolumn,twoside]{osajnl}
\usepackage{amsmath}
\usepackage{subcaption}
\journal{jocn} 

\setboolean{shortarticle}{false}

\title{Continuous Variable Based Quantum Communication in the Ocean}

\author[1,*]{Ramniwas Meena}
\author[2,+]{Subhashish Banerjee}

\affil[1,2]{Department of Physics, Indian Institute of Technology, Jodhpur, India-342030}

\affil[*]{meena.53@iitj.ac.in}
\affil[+]{subhashish@iitj.ac.in}

\begin{abstract}
Continuous Variable-Based Quantum Cryptography (CV-QKD) is an emerging field in quantum information science, offering unprecedented security for communication protocols by harnessing the principles of quantum mechanics. However, ocean environments pose unique challenges to quantum communication due to their distinct properties and characteristics. This work investigates the impact of turbulence on the transmission of Gaussian light beams used in a continuous variable-based quantum key distribution system for underwater quantum communication. The objective is to quantitatively analyze the induced losses and propose methodologies to mitigate their effects. To achieve this, we adopt the widely accepted ABCD matrix formalism, which provides a comprehensive framework for characterizing the propagation of optical beams through different media. Moreover, a numerical simulation framework is developed to assess the resulting losses and evaluate the performance of the proposed system. The implications of these numerical simulation frameworks for the design and optimization of quantum communication systems for oceanic environments are thoroughly discussed. 
\end{abstract}

\setboolean{displaycopyright}{false} 

\begin{document}

\maketitle

\section{Introduction}

The field of underwater optical wireless communication (UOWC) has gained significant attention due to its potential for obtaining high data rates compared to acoustic, radio, magnetic, or electric fields \cite{zeng2016survey, hoeher2021underwater, cossu2017full}. Lasers, LEDs, and small LED arrays are suitable photon emitters for underwater optical communications. Due to its excellent beam collimation, laser-based communication has a higher range and signal-to-noise ratio than conventional optical transmitters. Unlike LEDs, the phase may be employed for transmission and receiving on both ends. Additionally, bandwidth is much higher than other techniques. The latency is much lower than auditory transmission, therefore implementation may be small and lightweight. Underwater optical wireless communication (UOWC) is limited by distance, absorption, and intensity scattering. These restrictions depend on aquatic quality.
\\

Despite the increased deployment of underwater sensor networks (USNs) and the expanding body of research on the subject, there has been a lack of emphasis on the cyber security elements. In marine applications, secure communication is essential for monitoring harbors, ports, offshore oil rigs, and undersea pipelines. It is also necessary for border security. Secure communication aims to protect data confidentiality, integrity, and authentication.
The advent of quantum computing has led to the need for "quantum-secure" cryptography techniques \cite{bennett1984proc,shor2000simple,mayers2001unconditional}. Quantum cryptography offers a level of security that is not dependent on untested mathematical principles, but rather on the hard rules of physics. This makes it a promising solution for ensuring unconditional security in a range of naval activities \cite{loepp2006protecting}. Continuous Variable-Based Quantum Cryptography (CV-QKD) is a notable technique \cite{grosshans2002continuous, srikara2020continuous} within the realm of quantum key distribution (QKD) owing to its exploitation of continuous-variable quantum systems.
Considerable advancements have been achieved in the advancement of CV-QKD protocols \cite{pirandola2020advances,ralph1999continuous, diamanti2015distributing,laudenbach2018continuous}. However, the implementation of these systems in intricate settings, especially underwater communication channels, presents several problems. Similar to quantum communication in free space \cite{ghalaii2022quantum, heim2014atmospheric, sharma2019analysis, sharma2018analysis, meena2023analysing}, underwater free space quantum key distribution (QKD) likewise encounters two inevitable challenges: saltwater attenuation and turbulence.

The phenomenon of absorption and multiple scattering has a significant role in underwater environments \cite{giuliano2019underwater,gussen2016survey}. These variables give rise to both power decrease and temporal dispersion, commonly known as intersymbol-interference (ISI). Consequently, this leads to a decline in the performance of the link and constraints on the lengths over which transmission can occur, often limited to approximately 100 meters. This assertion is supported by multiple sources found in the existing body of literature (e.g., \cite{tian2017high,wang2019100,zhang2020towards,dai2021200,fei2022100}).

Ocean is commonly acknowledged for its notable feature of exhibiting relatively minimal attenuation in the blue and green light range, particularly within the transmission window spanning from $450$ to $550$ nanometers, as evidenced by earlier research (e.g., \cite{duntley1963light,gilbert1966underwater}). The distinctive characteristic of this feature has established the foundation for the progression of underwater optical communication (UWOC) technologies. In the course of our investigation, we come accross this particular facet. An optimization analysis is performed to examine the attenuation coefficient at different wavelengths and its relationship with the ocean profile at varying depths. The results of our study provide additional insights into this issue. Another important aspect that has a substantial impact on underwater optical communication's (UWOC) performance is turbulence, which further restricts the technology's wide application. Although laser beam propagation through atmospheric turbulence has been extensively investigated both theoretically and experimentally (see, for example, works \cite{borah2007estimation, dutta2023analysis, coronel2023characterization, wu2023experimental}), the field of laser beam propagation through complex oceanic turbulence is still relatively unexplored and requires further research. Here, "turbulence" refers to the random variations in the water's refractive index, or "n", which are mostly caused by changes in the ocean's salinity and temperature. These variations result in variations in the received intensity, which lead to beam spreading and fading. Consequently, these phenomena have a detrimental impact on the performance of UWOC systems (as observed in studies like \cite{tang2013temporal,andrews2005laser}).

The complex and erratic oceanic movements pose a significant challenge for conducting field tests on turbulence, making these experiments both costly and time-consuming. Consequently, there arises a need for a straightforward simulation platform that can predict system performance before embarking on the design, implementation, and evaluation of underwater optical communication (UWOC) systems.

Given the multitude of factors influencing underwater optical channels and the intricate nature of their analysis, the Monte Carlo (MC) method has emerged as a crucial tool for investigating UWOC channels, offering precise numerical solutions for photon packet behavior. Notably, previous works (e.g., \cite{zhang2020monte,enghiyad2022impulse}) have employed this method to simulate how turbulence affects the direction of photon propagation, relying on Snell's law. Others have utilized the Fournier-Forand (FF) function to model changes in the scattering phase function induced by turbulence \cite{geldard2020effects}. The channel model was further extended through a Monte Carlo simulation, examining turbulence-induced attenuation using the wide-range Prandtl/Schmidt number power spectrum \cite{xu2022improvement}.

{However, these Monte Carlo methods primarily focus on simulating idealized light rays, emphasizing light irradiance and trajectories while primarily considering the particle-like behavior of light. Experimental investigations have also been conducted in this domain, such as examining turbulence's effect on the Gaussian beam waist over a few-meters water path \cite{wu2023experimental} and studying the influence of turbulence on underwater communication in a natural environment \cite{ji2023analysis, Hou:12}. Moreover, the index of refraction structure constant was experimentally examined in clear water over a vertical transmission span of approximately 9 meters \cite{Nootz:16}, and an empirical equation for ocean's refractive index (which includes dissolved salts) was provided \cite{quan1995empirical}.}

In this paper, our focus centers on presenting a simplified physical simulation model based on the ABCD matrix approach \cite{kogelnik1965imaging,Kogelnik:65} for a turbulent UWOC. The ABCD matrix enables the analysis of various optical phenomena, including refraction, diffraction, and other effects relevant to turbulence-induced losses.  This approach offers a significantly reduced computational burden compared to methods grounded in computational fluid dynamics (CFD) \cite{tu2023computational} and radiative
transfer equation (RTE) \cite{rtn8}. The proposed model revolves around the interaction between a propagating laser beam and a turbulent medium, represented as successive turbulent cells with varying refractive indices ($n$). These variations in $n$ result from temperature fluctuations, changes in salinity, and pressure gradients. Furthermore, our model takes into account the influence of dynamic water flow on the propagation of laser beams.

In addition, a numerical simulation framework has been built in order to examine the impact of turbulence on the  CV-QKD system's performance and estimate the resulting losses. This framework allows for the assessment of system performance under different turbulence situations. In this study, we employed non-Gaussian postselection and virtual photon subtraction techniques \cite{huang2013performance, malpani2019lower,meena2023characterization, malpani2020impact} in the context of CV-QKD \cite{li2016non} to evaluate the key rate for underwater optical wireless communication (UOWC).  The use of statistical models is essential in considering realistic situations that account for fluctuations in turbulence. The analysis focuses on quantifying performance measures, including channel transmittance and secure key rate, to assess the impact of turbulence on the performance of the system.

The present paper is structured in the following manner: In Section \ref{sec:channel}, we present an overview of the attenuation phenomenon and propose a turbulence model to describe the propagation characteristics of a Gaussian laser beam. In Section \ref{sec:protocol}, we present an overview of the fundamental concepts behind photon subtraction and the corresponding postselection technique employed in a coherent-state continuous-variable quantum key distribution (CV-QKD) protocol. In Section \ref{sec:results}, we will discuss the numerical simulation and results for ocean channel parameters and explore their influence on keyrate. The summary is provided in Section \ref{sec:conclusion}.

\section{Underwater Channel} \label{sec:channel}

Light waves can travel long distances underwater, but they are subject to absorption and scattering by water particles and impurities \cite{giuliano2019underwater,gussen2016survey}. This absorption can limit the range and effectiveness of the communication.  Water turbidity and suspended particles can scatter the laser beam, reducing its strength and causing signal loss. High turbidity levels and long transmission distances can be particularly challenging for underwater laser communication.

The absorption and scattering is briefly discussed in part A, while for the turbulence we propose a model in part B.

\subsection{Absorption and Scattering Models}
Underwater, light attenuation occurs due to two main factors: scattering and absorption. Scattering causes the redirection of electromagnetic radiation away from its original path, while absorption converts the energy of the radiation into other forms, like heat or chemical energy. Both of these processes are influenced by the wavelength of light, denoted as $\lambda$ (measured in $nm$). We can express the total light attenuation as the sum of absorption $(a)$ and scattering $(b)$ coefficients, both measured in units of $m^{-1}$ (for depth d inside the ocean) \cite{xiang2021improving}:
\begin{equation}
    c(\lambda, d)=a(\lambda,d )+b(\lambda,d).
    \label{Eq:attenuation}
\end{equation}

The absorption and scattering spectra in oean are shaped by distinct biological factors, categorized based on their optical behaviors. These factors include the absorption characteristics of pure water, the absorption attributed to chlorophyll-a, which constitutes a significant portion of phytoplankton - a collection of photosynthesizing microorganisms, and the absorption connected to humic and fulvic acids, both of which serve as nutrients essential for the growth of phytoplankton. Each of these biological elements plays a crucial role in determining the optical properties of ocean, influencing its overall light absorption and scattering behaviors.\par

\begin{figure}[ht]
    \centering
    \includegraphics[width=0.45\textwidth]{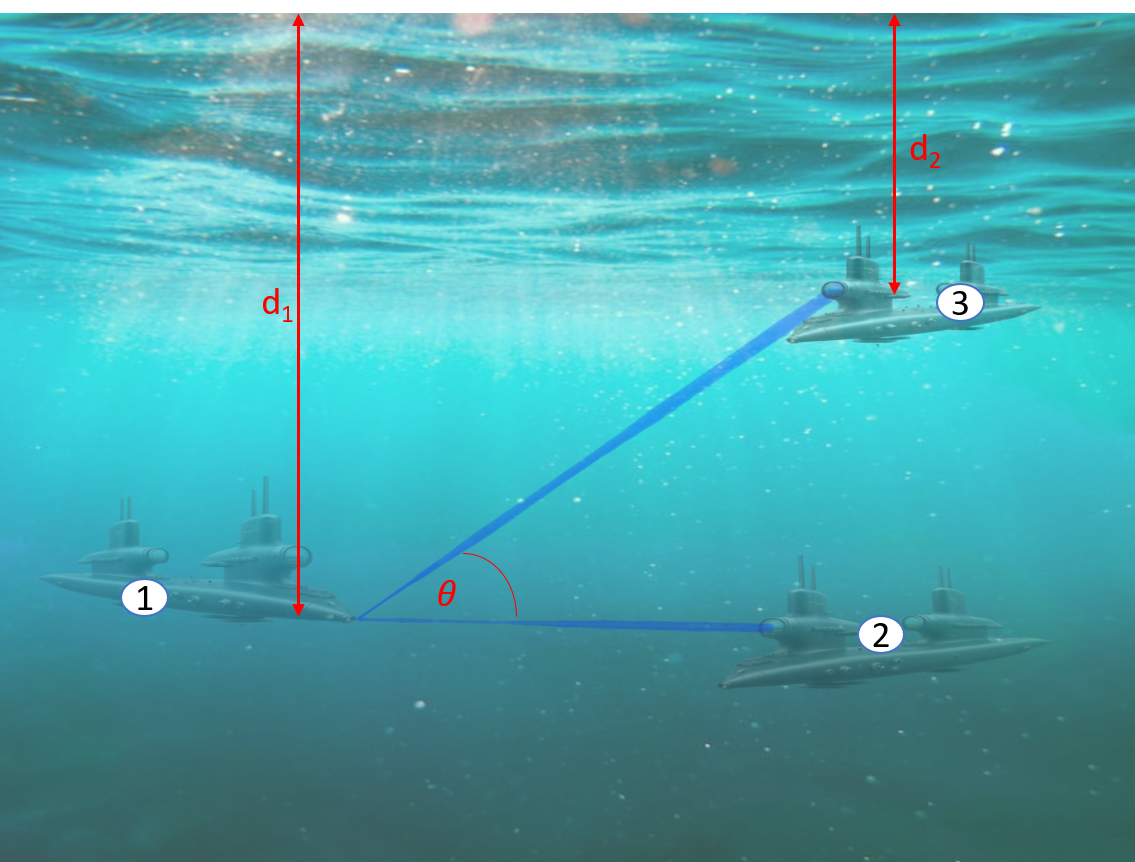}
    \caption{Underwater Communication Scenarios for Nuclear Submarines. There are two situation occur, one when  both legitimate parties (Nuclear Submarine 1 and 2) at same level and second when both legitimate parties at different depth (Nuclear Submarine 1 and 3).}
    \label{Nuclear_submarine}
\end{figure}

 Figure \ref{Nuclear_submarine} illustrates two distinct communication scenarios involving nuclear submarines:

Same Depth Communication (SDC) : In first scenario, both legitimate parties, represented by Nuclear Submarine 1 and Nuclear Submarine 2, are positioned at the same depth level, facilitating direct and efficient communication.

Different Depth Communication (DDC) : In second scenario, the legitimate parties (Nuclear Submarine 1 and Nuclear Submarine 3) are located at different depths within the ocean. This situation introduces challenges and considerations for effective underwater communication between submarines operating at varying depths.

Under the influence of attenuation loss, the intensity of transmitted signals follows Beer's Law. However, it's important to note that the attenuation coefficient varies with depth, leading to variations in intensity for communication scenarios involving the same depth (Same Depth Communication) and those with different depths (Different Depth Communication). The intensity at receiver end can be written as follow:

\begin{equation}
    I_a(d,z)=I_0 \times \bar{I}_{d,z}
\end{equation}
where,
\begin{equation*}
    \bar{I}_{d,z}=\begin{bmatrix}
        e^{-c (\lambda,d)\;z} && \text{for SDC}\\
        e^{-\int{c (\lambda,x)\;dx/\sin{\theta}}} && \text{for DDC}  
    \end{bmatrix}.
\end{equation*}

\subsection{Turbulence}
In underwater optical communication (UOWC), apart from challenges like scattering and absorption, there's a significant issue known as underwater optical turbulence (UOT). To understand UOT better, it's important to note that the refractive index of water (denoted as $n$) depends on factors like temperature, salinity, and wavelength. While in relatively stable and small water volumes, these factors can be assumed constant, in environments like seas and oceans, refractive index varies with time and location. This variability is due to the complex interaction of ocean currents, ranging from large-scale currents spanning thousands of kilometers to smaller-scale eddies down to centimeter scales. 

Large-scale ocean currents give rise to mesoscale eddies, typically spanning 10 to 100 kilometers. These mesoscale eddies, in turn, interact and create submesoscale turbulent cells, ranging from 100 meters to 10 kilometers. The motions in these scales are primarily horizontal (two-dimensional). Finally, submesoscale eddies generate microscale turbulence, which operates on scales from a few centimeters to 100 meters. Importantly, turbulence is three-dimensional only below the 100-meter mark and is referred to as microscale turbulence.

Microscale turbulence in the upper ocean is caused by factors such as surface winds and evaporation. Within the ocean, microscale turbulence emerges when internal waves generate strong shears, leading to overturning and breaking. These internal microscales result from the mixing of dense and light waters.
One commonly used empirical equation for the refractive index of ocean (containing dissolved salts), which provides an approximation of the refractive index as a function of temperature, pressure, and salinity was provided in \cite{quan1995empirical,mcneil1977metrical} and is commonly used in oceanography and underwater optics.

\begin{multline}
    n(S,T,\lambda)=n_0+\left(n_1+n_2 T+n_3 T^2\right)S\\
    +n_4 T^2+\frac{n_5+n_6 S+n_7T}{\lambda}+\frac{n_8}{\lambda^2}+\frac{n_9}{\lambda^3}, \label{emparical}
\end{multline}
where S is the salinity in $\%$, T is the temperature in degree Celsius and $\lambda$ is the wavelength in nm. The ranges of validity are $0^{o} < T < 30^{o} C$, $0\% < S < 35\%$, and $400 nm < \lambda < 700 nm$.
\begin{table}[ht]
    \begin{tabular}{|c|c|}
    \hline
        $n_0=1.31405$ &  $n_1=1.1779 \times 10^{-4} $\\
        \hline
        $n_2=-1.05\times 10^{-6}$ & $n_3=1.6\times 10^{-8}$\\
        \hline
        $n_4=-2.02\times 10^{-6} $& $n_5=15.868$\\
        \hline
        $n_6=0.01155$  & $n_7=-0.00423$\\
        \hline
        $n_8=4328$ & $n_9=1.1455\times 10^{6}$\\
        \hline
    \end{tabular}
    \caption{Coefficients in \eqref{emparical} with Corresponding Values.}
    \label{refractive_index_parameter}
\end{table}

\subsubsection*{Discrete cell Turbulence Model (Our Proposed Model)}
We design a model to estimate turbulence inside the ocean for point to point link. The model involves using a series of random curvatures and random distances between them. To describe the propagation with randomly varying refractive index ($\delta n$) and  curvature $R$  of surface separating two cells, the ABCD approach can be used. This method simplifies the calculation of the beam's transformation through optical elements and media. We make the following assumptions to implement the model:

1. Propagation Medium: The medium in the $x$ and $y$-axis directions is assumed to be unlimited. This implies that there are no constraints on the propagation of light in the horizontal plane.

2. Boundary Interaction: The interaction of light with the water-air interface is ignored. In other words, the light is assumed to propagate freely without any reflection or refraction inside the water medium.

3. Absorption and Scattering: The model simplifies the simulation by neglecting absorption and scattering phenomena.

\begin{figure*}[ht]
    \centering
    \includegraphics[width=0.8\linewidth]{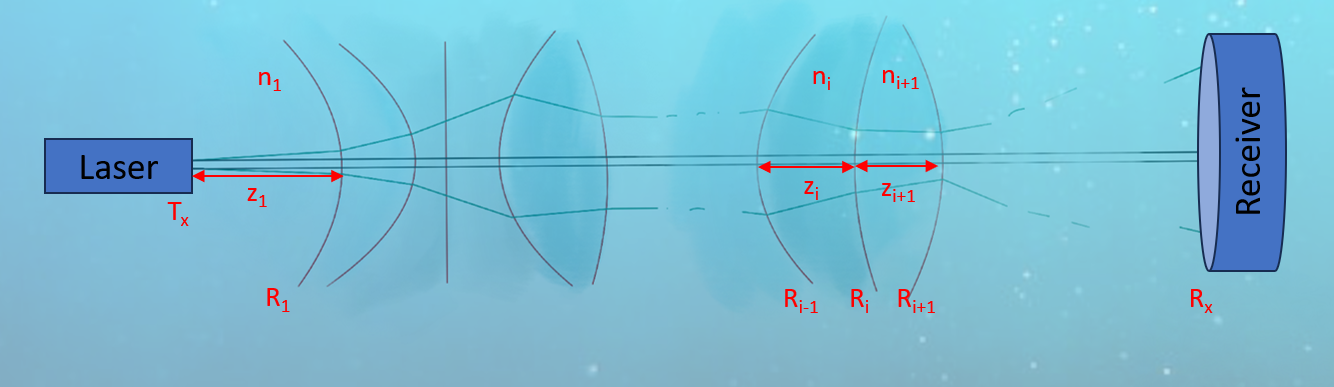}
    \caption{Schematic illustration of the beam propagation model in the Underwater Optical Communication (UWOC) channel. The parameter \(z\) denotes the distance between two adjacent cells, and \(R\) represents the radius of curvature of the cell's surface. \(T_x\) indicates the transmitter aperture radius, while \(R_x\) corresponds to the receiver aperture radius. The green line traces the path of a Gaussian beam as it passes through the turbulent cells.}
    \label{fig:turb_model}
\end{figure*}

4. Turbulence cells: The space between the transmitter ($T_x$) and receiver ($R_x$) is divided into multiple turbulence cells ($k$ cells), refer to Fig. \ref{fig:turb_model}. Each turbulence cell is separated by curved borders.

5. The Gaussian beam is well collimated, allowing us to apply Gaussian beam optics.

The ABCD matrix describe the transformation of a Gaussian beam through an optical system or medium using four parameters: A, B, C and D. These parameters are related to the optical properties of the system and can be used to calculate the beam's properties at different points along the path.
Here, we consider a scenario where the turbulent water medium is divided into series of discrete cells, each with a random curvature ($R_i$) and separated by random distances ($z_i$).
By cascading the ABCD matrices of all turbulence cells, we can obtain the overall ABCD matrix of the entire path between the $T_x$ and $R_x$. This will give us information about how the light beam’s parameters change throughout the propagation due to turbulence.

Let's break down the steps:

1. As previous described, we have the initial Gaussian beam parameters:
Waist radius ($w_0$), 
waist position ($z_0$), 
wave number ($\Bar{k}=2\pi/\lambda$), 
complex parameter $q_0=z_0+\iota \left(\bar{k} w_0^2/2\right)$.

2. Propagation through each cell:
For each cell, we will calculate the ABCD matrix based in the random curvature $R$ and propagation distance $z_i$. The $ABCD$ matrix for $i^{th}$ cell will be:
\begin{align}
    P_i = \begin{bmatrix}
    1 & z_i/n_i \\
    0 & 1 \\
\end{bmatrix}
\end{align}

\begin{align}
    C_i = \begin{bmatrix}
    1 & 0 \\
    (n_{i+1}-n_i)R_i/n_i & n_{i+1}/n_i \\
\end{bmatrix}
\end{align}
where, $P_i$ for propagation through $i^{th}$ cell and $C_i$ for propagation from $i^{th}$ Curvature.

3. Cumulative ABCD Matrix for Full Propagation:
To account for the propagtion through multiple cells, we need to accumulate the ABCD matrices for each cell. we will start with initial complex parameter $q_0$ and sequentially apply the ABCD matrices of each cell 
\begin{equation}
    ABCD=\Pi^{k}_{i=1}P_i*C_i.
\end{equation}

4. Final Gaussian Beam Parameter (output):
After propagating through all cells, we can calculate the final Gaussian beam parameters using the follow at waist position z:
\begin{equation}
    q_{out}=\frac{A q_{0}+B}{C q_{0}+D}.
\end{equation}

The complex beam parameter at the beam position \(z\), denoted as \(q_{out}\), is given by \(q_{out} = z + i\left(\bar{k} w^2(z)/2\right)\), where \(i\) is the imaginary unit. From this parameter, we can extract the beam waist \(w(z)\), allowing us to determine the irradiance distribution of a Gaussian \(TEM_{00}\) beam \cite{griffiths2005introduction},
\begin{equation}
    I_{turb}(r,z)=I_{0} e^{-2r^2/w^2(z)}.
\end{equation}

The discrete-cell model presented herein provides a structured approach to analyze the effects of turbulence on Gaussian beam propagation through a turbulent water medium. By incorporating the ABCD matrix formalism and discrete-cell representation, this study contributes to a deeper understanding of how random curvatures and spatial separations within turbulent media impact the behavior of optical beams. It is important to note that while this model offers valuable insights, it remains a simplified representation of real-world turbulence, and more advanced numerical techniques may be required for a comprehensive analysis of complex turbulent environments.

After considering the absorption, scattering and turbulence, the irradiance distribution of Gaussian  $TEM_{00}$ beam at waist position $z$ depth and depth $d$ can be written as follow:-
\begin{equation}
    I(r,d,z)=I_{turb}(r,z) \times  \bar{I}(d,z)\label{intensity_total}.
\end{equation}

The total channel transmittance in an underwater communication channel is defined as the ratio of the received power (\(P_{\text{received}}\)) at the receiver to the transmitted power (\(P_{\text{transmitted}}\)) at the transmitter. Mathematically, it can be represented as:
\begin{equation}
    \text{Channel Transmittance}~ (T_c) = \frac{P_{\text{received}}}{P_{\text{transmitted}}} \label{channel_transmittance},
\end{equation}

where,
\(P_{\text{received}}=\int^{2\pi}_0\int^{R_x}_0 {I(r,d,z)rdrd\theta}\) is the power measured at the receiver, representing the signal strength after it has propagated through the underwater channel and \(P_{\text{transmitted}}=\int^{2\pi}_0\int^{T_x}_0 {I(r,d,0)rdrd\theta}\) is the power initially emitted by the transmitter, denoting the signal strength at the point of transmission. This channel transmittance ratio provides a fundamental measure of how efficiently the transmitted power is conveyed through the underwater medium and is a critical parameter in the analysis and characterization of underwater communication systems.

\section{Protocol and the secure key rate analysis} \label{sec:protocol}
First, let's start by introducing the basics of virtual photon subtraction in the context of Alice's laboratory, as depicted in Fig. \ref{fig:QKD_setup}. We begin by applying photon subtraction to the two-mode squeezed vacuum (TMSV) source, characterized by a squeezing parameter denoted as $\zeta$ \cite{li2016non}. After passing through the beam splitter, mode B splits into two modes, namely $B_1$ and $B_2$. This setup results in Alice having a tripartite state represented as $\rho_{AB_1B_2}$.

In the next step, we perform measurements on mode $B_1$ using a Positive Operator-Valued Measure (POVM) with two possible outcomes: $\hat{\Pi}_1$ and $\hat{\Pi}_2$ (or $I - \hat{\Pi}_1$). The modes $A$ and $B_2$ are retained only when the POVM measurement yields a click after the $\hat{\Pi}_1$ operation.

Different choices of the POVM element $\hat{\Pi}_1$ correspond to various methods of photon subtraction. In this discussion, our primary focus is on the specific process of subtracting $m$ photons, which is represented by $\hat{\Pi}_1 = |m\rangle\langle{m}|$.

Following Alice's heterodyne measurement on mode A and her POVM measurement (represented by operators $\hat{\Pi}_1$ and $\hat{\Pi}_2$) performed on mode $B_1$, it is important to note that these measurements commute. This means that the state's knowledge will collapse the other mode onto a coherent state. After the beam splitter $BS_1$, the mathematical description of the state for modes $B_1$ and $B_2$, given Alice's measurement results $x_A$ and $p_A$, can be expressed as $|\phi^{(x_A,p_A)}\rangle_{B_1B_2} = |\sqrt{1-T}\alpha\rangle_{B_1}|\sqrt{T}\alpha\rangle_{B_2}$, where $\alpha=\sqrt{2} \zeta (x_A-\iota p_A)/2$.

The probability of successfully subtracting $m$-photons is then determined by the values of {$x_A,p_A$}, and is given by:
\begin{align}
    & P^{\hat{\Pi}_1(m|x_A, p_A)} = {|\langle m|\sqrt{1 - T\alpha}\rangle|}^2 \nonumber \\
    &= \exp\left[-\frac{{(1-T)\zeta^2}}{2}(x_A^2 + p_A^2)\right]\cdot\left[\frac{{(1-T)\zeta^2}}{2}(x_A^2 + p_A^2)\right]^m / m!.
\end{align}

Now, let's consider the mixed state that emerges from Alice's setup, denoted as $\rho^{(m)}_{B_2}$. This state is determined by integrating over all possible values of {$x_A, p_A$} and applying a specific weighting function:

\begin{equation}
    \rho^{(m)}_{B_2} = \int dx_A dp_A W P_{x_A, p_A} | \sqrt{T\alpha} \rangle \langle \sqrt{T\alpha} |
\end{equation}

In this equation, the weighting function $W = {{P}^{\hat{\Pi}_1}(m|x_A, p_A)}/{{P}^{\hat{\Pi}_1}(m)}$ accounts for the probabilities associated with different values of $\{x_A, p_A\}$ in the integration process and $P(x_A, p_A) = {1}/{(\pi(V + 1))} \exp\left(-\frac{x_A^2 + p_A^2}{V + 1}\right)$ is the Gaussian distribution of Alice's heterodyne measurement results, and $V =(1 + \zeta^2 )/(1-\zeta^2 )$ is the variance of the TMSV state.
In comparison to the scenario where Alice doesn't perform any photon subtraction operations, resulting in an output mixed state given by:

\begin{equation}
\rho^{(G)}_B=\int{dx_{A} dp_{A} P(x_A,p_A)\left|\alpha\rangle\langle\alpha\right|}
\end{equation}
Let's clarify the differences between the scenario where Alice performs photon subtraction operations and the scenario where she doesn't, resulting in output mixed states:

In the photon-subtraction scenario, there's an additional factor known as the weighting function, denoted as $W$, which serves as a filter or acceptance probability for different pairs of $\{x_A, p_A\}$. It represents the probability of accepting a specific pair of {$x_A, p_A$} given the photon subtraction operation. This filter function can be expressed as:
\begin{equation}
    Q\left(\gamma, \zeta, T\right) = {P}^{\hat{\Pi}_1}(m)W ={P}^{\hat{\Pi}_1}(m|x_A, p_A) \label{accepted_probability}
\end{equation}

The second difference is that in the photon-subtraction scenario, the output coherent state needs to undergo a beam splitter operation with transmittance $T$. This operation effectively scales down the mean value of the output coherent state to $\sqrt{T}\alpha$.

In summary, by changing the perspective on the protocol and exchanging Alice's two commutable measurements, we achieve an equivalent virtual photon subtraction by postselecting Alice's heterodyne measurement results and scaling the mean value of the output coherent state by a factor of $\sqrt{T}$. The postselection filter function is represented by equation \eqref{accepted_probability}.  

\subsection{Protocol}
\begin{figure}[ht]
    \centering
    \includegraphics[width=0.9\linewidth]{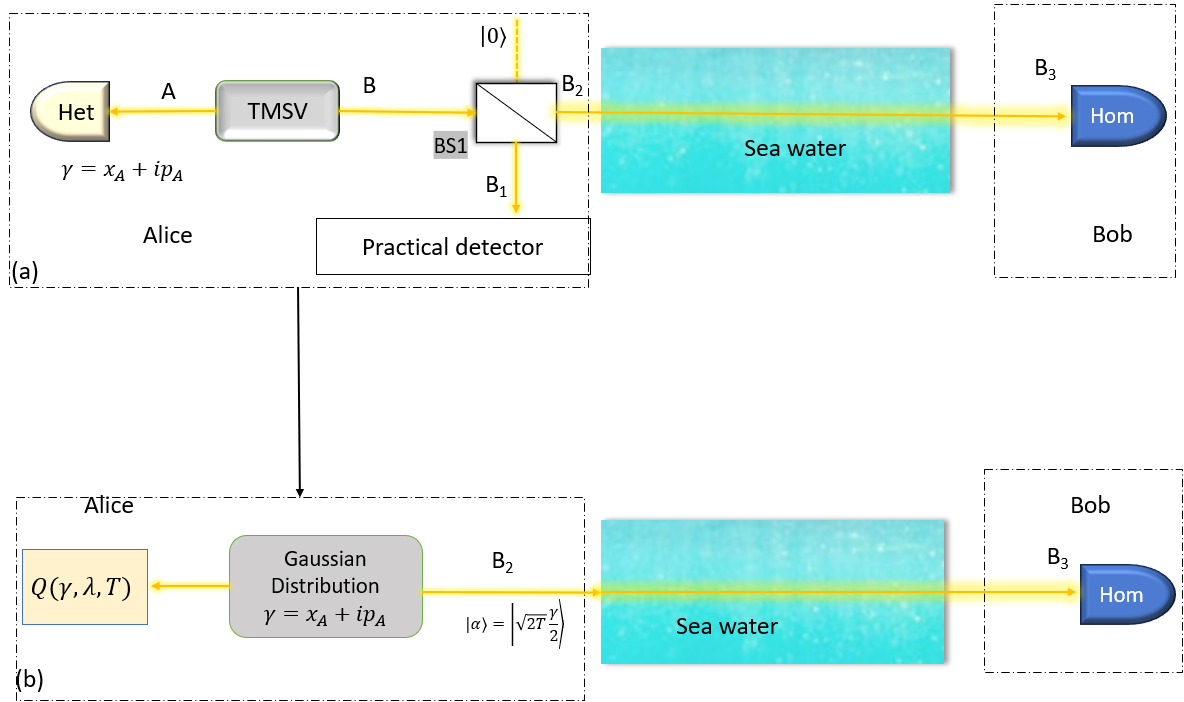}
    \caption{Illustration of (a) the Entanglement-based (EB) scheme of CV-QKD with photon subtraction and (b) the Prepare-and-measure (PM) scheme of CV QKD with equivalent postselection as virtual photon subtraction. Het: heterodyne detection; Hom: homodyne detection; BS1: beam splitter; $\gamma$: Alice’s measurement result; $\zeta$: parameter of two-mode squeezed vacuum (TMSV); $Q(\gamma ,\zeta,T)$: postselection filter function; $T(\eta)$: transmittance of BS1; $T_c ,\epsilon$: channel parameters.}
\label{fig:QKD_setup}

\end{figure}

Here, we are present a concise description of the Photon-Subtraction Measurement (PM) scheme designed for CV-QKD, as depicted in the accompanying  Fig. \ref{fig:QKD_setup}. The protocol unfolds as follows:

Step 1: Alice initiates the process by generating a coherent state denoted as $|\alpha \rangle $, where $\alpha =\sqrt{2 T}\gamma /2 $, with $\gamma =x_A+\iota p_A$. Here, $\{x_A, p_A\}$ are chosen randomly from a Gaussian-distributed set with a mean of zero and variances $V_{x_A} = V_{x_B} = (V + 1)/2$. The parameter $T$ represents the transmittance of the beamsplitter $BS_1$, and $V$ signifies the variance associated with the equivalent Two-Mode Squeezed Vacuum (TMSV) state. Alice then transmits this coherent state to Bob.

Step 2: Upon receiving the transmitted state, Bob proceeds to perform either homodyne or heterodyne detection, yielding measurement outcomes denoted as $x_B$ and $p_B$.

Step 3: Steps 1 and 2 are iteratively executed until a sufficient amount of data is collected. It is Alice's responsibility to determine which data points are acceptable, and she communicates these decisions to Bob. The probability of accepting each data point is determined by $Q(\gamma,\zeta, T)$ as defined in \eqref{accepted_probability}.

Subsequently, Alice and Bob utilize the accepted data to progress through post-processing steps, including parameter estimation, information reconciliation, and privacy amplification. Notably, Alice discloses her decision regarding the acceptance or rejection of each data point after Bob's measurements. The discarded data points can be regarded as decoy states, a concept reminiscent of a previous non-Gaussian protocol \cite{leverrier2011continuous}.

As this protocol exclusively involves the Gaussian state, it's sufficient to rely on the covariance matrix, denoted as $\gamma^{(m)}_{AB_3}$, to calculate the rate. This rate can be computed based on the accepted data from both Alice and Bob when practically implementing this protocol.

In our simulation, we make the assumption that the channel characteristics can be described using two parameters: the channel transmittance, denoted as $T_c$, and the excess noise, represented as $\epsilon$. If we have the covariance matrix of $\gamma^{(m)}_{AB_2}$, it can be expressed as follows:

\begin{equation}
\gamma^{(m)}_{AB_2} = \begin{bmatrix}
V_A I_2 & \phi_{AB}\sigma_z \\
\phi_{AB}\sigma_z & V_B I_2
\end{bmatrix}
\end{equation}

where, \(V_A\) and \(V_B\) represent the variances in modes \(A\) and \(B_2\), respectively. The covariance between the quadratures of modes \(A\) and \(B_2\) is denoted as \(\phi_{AB}\). The matrices \(I_2\) and \(\sigma_z\) are defined as diagonal matrices with elements \((1,1)\) and \((1,-1)\), respectively.
Then after the channel transmittance,
   \begin{equation}
     \gamma^{(m)}_{AB_3}=\begin{bmatrix}
         V_A I_2 & \sqrt{T_c}\phi_{AB}\sigma_z\\
         \sqrt{T_c}\phi_{AB}\sigma_z & T_c(V_B+\chi) I_2
     \end{bmatrix}.
 \end{equation}
We introduce the parameter $\chi$, defined as $\chi=(1-T_c)/T_c+\epsilon$, where $T_c$ represents the channel transmittance, and $\epsilon$ represents access noise (any external interference or disturbance that impacts the light beam being transmitted or detected). The detailed expression for $\gamma^{(m)}_{AB_2}$ is provided in the Appendix C. Throughout the remainder of this paper, we make the assumption that Bob employs homodyne detection.

\subsection{Keyrate}
 Usually, the secret key rate of the TMSV state is no less than the secret key rate of the equivalent Gaussian state, which shares an identical covariance matrix due to the extremality of Gaussian state \cite{navascues2006optimality, garcia2006unconditional, wolf2006extremality}. Hence, we will use $\gamma^{G}_{AB_2}$ to derive the lower bound of the secret key state. Beside the acceptance probability for each of the data in the post-selection step should also be taken into account. This probability is equivalent to the success probability of Alice's POVM measurement $P^{\hat{\Pi}_1}$ and can be treated as a scaling factor. 

We analyze the performance of the underwater CV-QKD system based on the parameters discussed above. The theoretical asymptotic secret key rate under collective attack is obtained by 
\begin{equation}
    K^{Hom}=P^{\hat{\Pi_1}}(m)\left(\beta I^{Hom}_{AB}-\chi^{Hom}_{BE}\right),
    \label{eq:keyrate}
\end{equation}
 where $\beta$ is the reverse reconciliation efficiency, $I^{Hom}_{AB}$ is the Shannon mutual information between Alice and Bob, and $\chi^{Hom}_{BE}$ is the Holevo quantity which bounds the maximum information available to Eve and the superscript  Hom means Bob using homodyne detection. The detailed expressions for these quantities are provided in Appendix C. \par

\section{Results and Discussion} \label{sec:results}

Our investigation encompasses three key areas:  turbulence modeling, channel transmittance analysis and the application of CV-QKD in underwater communication. The results are now discussed.

\begin{figure}[ht]
  \centering

  \begin{subfigure}{0.48\textwidth}
  \centering
    \includegraphics[width=\linewidth]{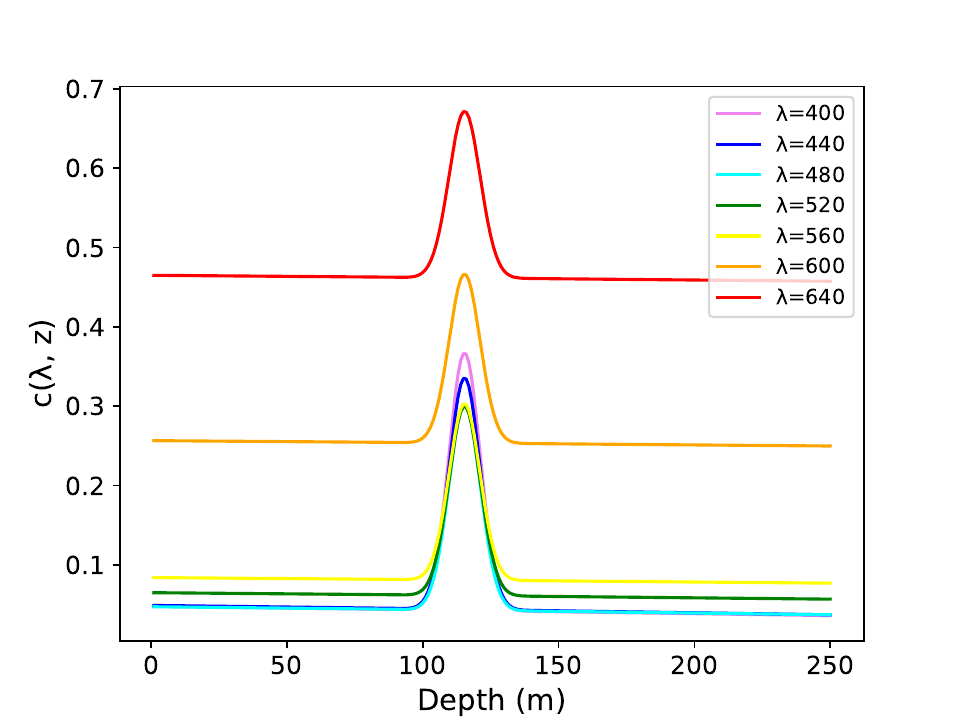}
    \captionsetup{justification=centering} 
    \caption{}
    \label{fig:Attenuation_wavelengthS1}
  \end{subfigure}
  \hfill
  \begin{subfigure}{0.48\textwidth}
  \centering
    \includegraphics[width=\linewidth]{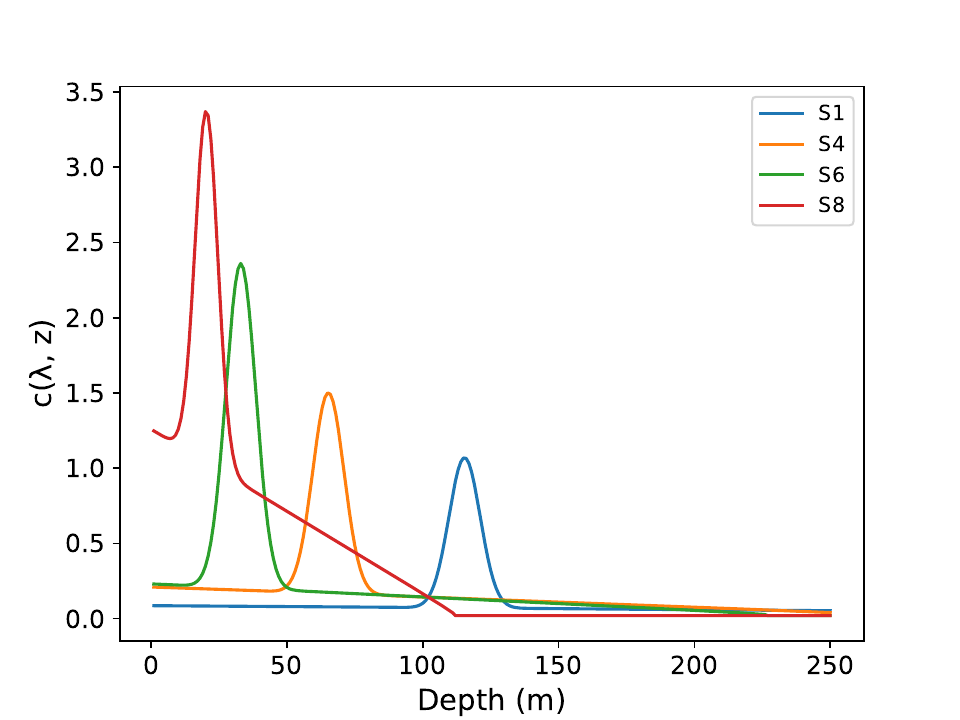}
    \captionsetup{justification=centering} 
    \caption{}
    \label{fig:Attenuation_S1_8}
  \end{subfigure}

  \caption{(a) Attenuation Coefficient as a Function of Depth (from ocean Surface) for Different Wavelengths ($S_1$ type water), (b) Attenuation Coefficient vs. Depth from ocean Surface for Various Types of ocean ($\lambda =480nm$).}
  \label{fig:Attenuation_seawatertype}
\end{figure}

To optimize the selection of wavelength and ocean depth based on ocean water types for channel analysis and quantum communication, we have plotted attenuation coefficient  with depth of ocean (utilizing \eqref{Eq:attenuation} for different wavelengths) and different ocean water profiles, and are depicted in Figs. \ref{fig:Attenuation_wavelengthS1} and \ref{fig:Attenuation_S1_8}, respectively. Notably, each water profile exhibits a peak at a specific depth, denoted as $d_{max}$. This depth corresponds to the chlorophyll concentration maximum for the respective water profile. At $d_{max}$, the chlorophyll concentrations reach their saturation point causing a distinctive peak in attenuation coefficient.
In Fig. \ref{fig:Attenuation_wavelengthS1}, a distinct trend is observed: as we transition wavelengths from approximately 400 to 480 nm, the attenuation coefficient decreases. Beyond this range, it begins to increase uniformly across all depths. In Fig. \ref{fig:Attenuation_S1_8}, the maximum attenuation values, occurring at $d_{max}$, decrease as we move towards $d_{\infty}$, where the effect of chlorophyll is negligible. Beyond $d_{\infty}$, only the impact of pure water is considered, as the chlorophyll concentration is nearly zero (see Table \ref{table:water_profile} for the relation between chlorophyll concentration at ocean surface, $d_{\infty}$ and $d_{max}$ are given for different water profiles). To enhance quantum communication efficiency, we strategically position our equipment below the $d_{max}$ depth, and we specifically opt for the $480 nm$ wavelength to minimize attenuation.

\begin{figure*}[ht]
    \centering
    \includegraphics[width=0.9\linewidth]{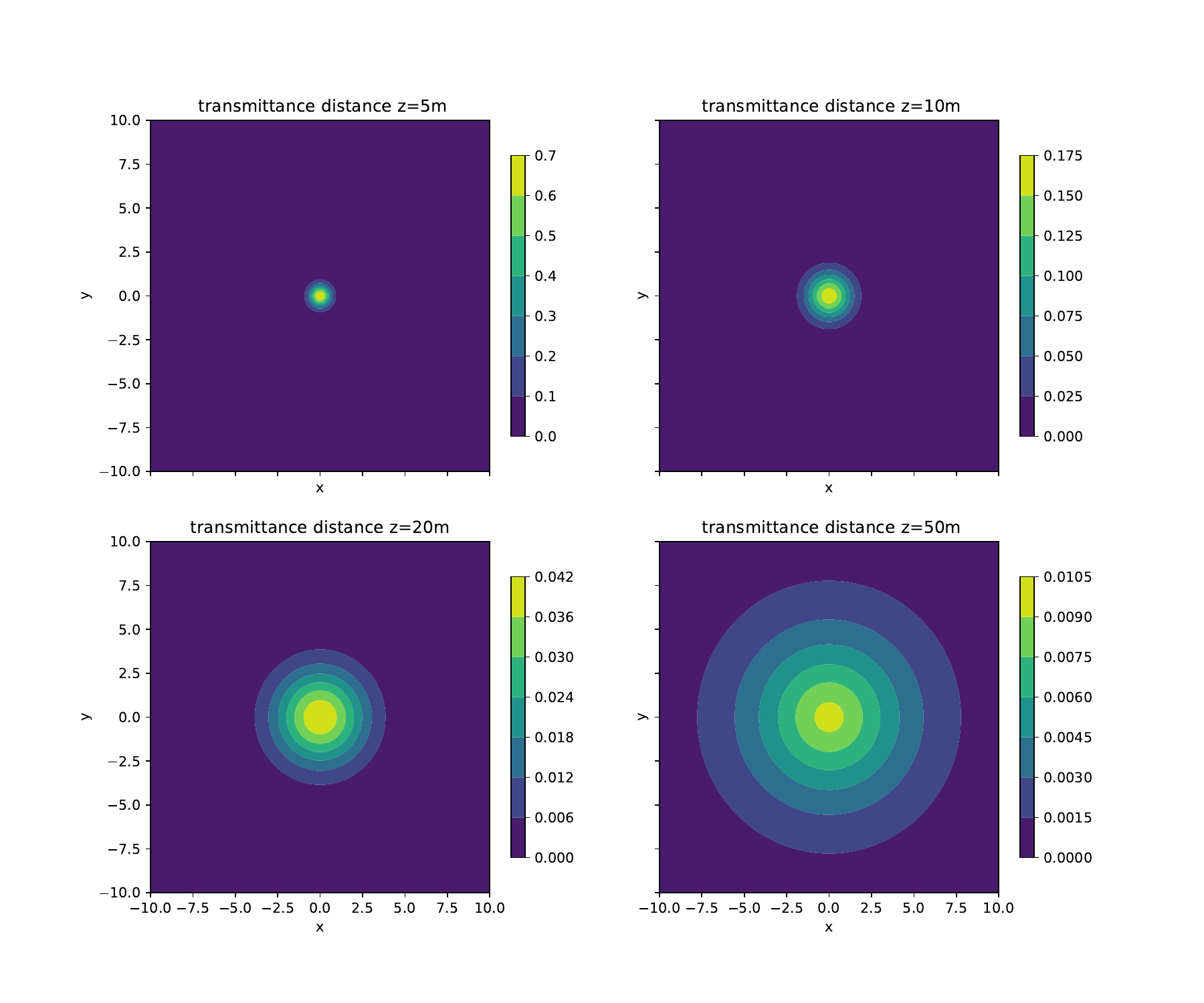}
    \captionsetup{justification=centering} 
    \caption{ Intensity distribution at received spots for transmittance distances $z = 5m,\;10m,\;20m $ and $50m$ (due to turbulence only).}
    \label{fig:Intensity_profile_turbulance}
\end{figure*}

\begin{figure}[ht]
    \centering
        \begin{subfigure}{0.49\textwidth}
        \centering
        \includegraphics[width=\linewidth]{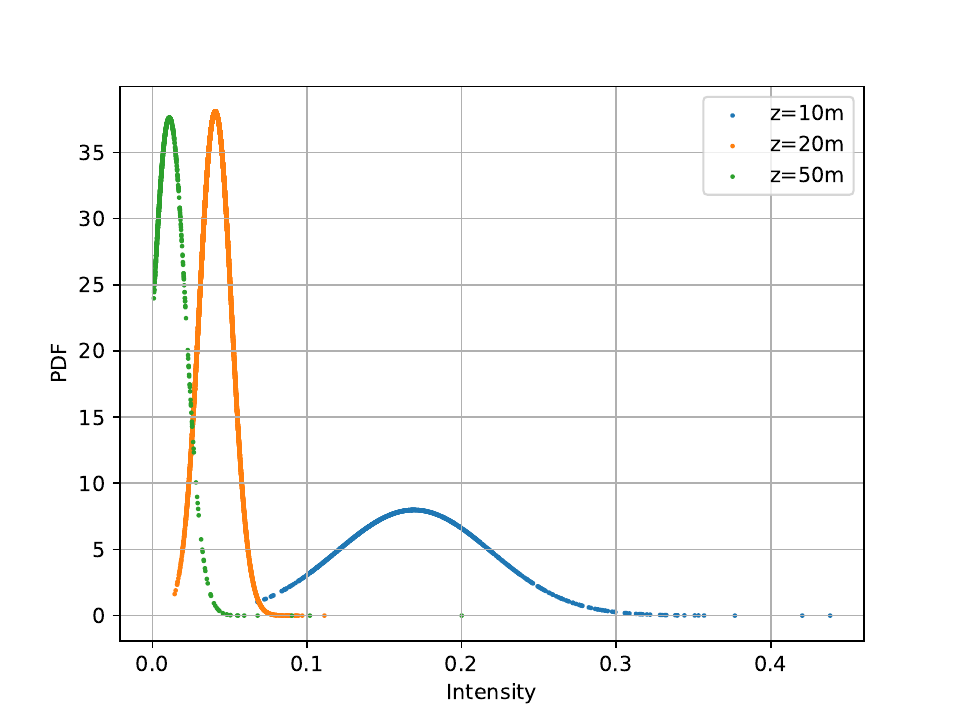}
        \caption{PDF with different transmittance distance for $5 cells/m$.}
        \label{fig:PDF_distance}
    \end{subfigure}
    \hfill
    \begin{subfigure}{0.49\textwidth}
        \centering
        \includegraphics[width=\linewidth]{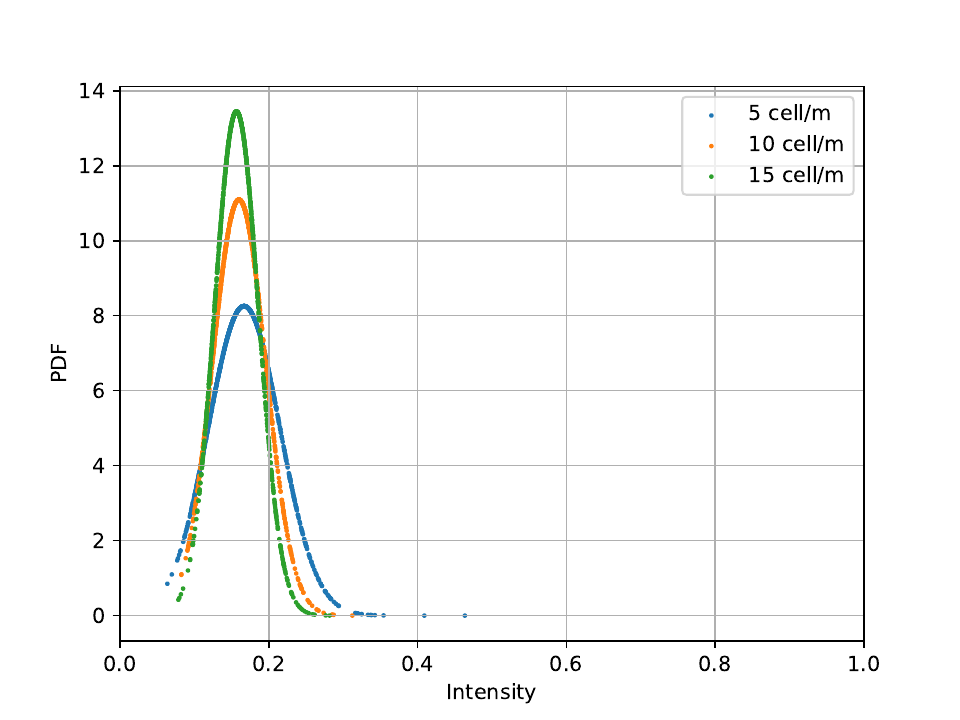}
        \caption{PDF with different turbulence cells at transmittance distance $10m$.}
        \label{fig:pdf_cell}
    \end{subfigure}

    \caption{PDF with received Intensity (only considering turbulence loss). }
    \label{fig:PDF}
\end{figure}

Our turbulence model was constructed with careful consideration of parameters critical to its accuracy. These parameters included a Gaussian beam with a waist size ($w_0$) of 0.04 m, transmitter and receiver aperture radii ($T_x$ and $R_x$) set at 0.08 m and 0.8 m, respectively. Additional factors such as water salinity ($12.5\%$), water temperature (290 K), and an average of 5 cells per meter were  included. To visually represent our turbulence model, we created contour plots illustrating the average beam intensity profile at different transmittance distances (see Fig. \ref{fig:Intensity_profile_turbulance}). As the transmittance distance ($z$) increases, the beam profile broadens, revealing the evolving characteristics of the turbulence in our model. To analyse the turbulence strength with different transmittance distance and average cells/m, the probability density function (PDF) with normalized received intensity ($I/I_0$)  has been  plotted in Fig. \ref{fig:PDF}. Here, the $PDF(I) = {1}/{\sqrt{2\pi\sigma^2}} \exp(-{(I - \mu)^2}/{2\sigma^2}$ , where \(I\) represents the received intensity, \(\mu\) is the mean value of the received intensity, \(\sigma\) is the standard deviation. An increased dispersion in the distribution indicates more pronounced fluctuations in light over transmittance distance and stronger turbulence. In Fig. \ref{fig:PDF_distance}, the illustration highlights that as the transmission distance increases, there is a noticeable reduction in fluctuations. Simultaneously, when examining the figure illustrating the relationship between the average number of turbulence cells per meter (as seen in Fig. \ref{fig:pdf_cell}), it is observed that an increase in this parameter results in a decrease in fluctuations, accompanied by a slight concurrent rise in turbulence at the same transmittance distance.

\begin{figure*}[ht]  
    \centering
    \begin{subfigure}{0.48\textwidth}
        \centering
        \includegraphics[width=\linewidth]{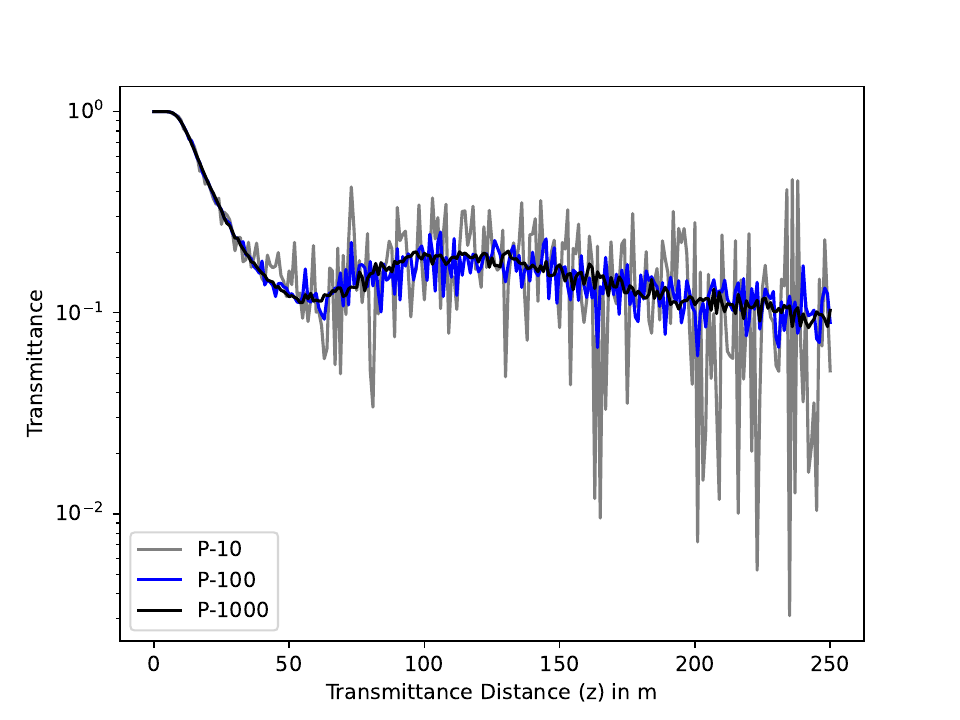}
        \captionsetup{justification=centering} 
        \caption{}
        \label{T-10-1000.png}
    \end{subfigure}
    \hfill
    \begin{subfigure}{0.48\textwidth}
        \centering
        \includegraphics[width=\linewidth]{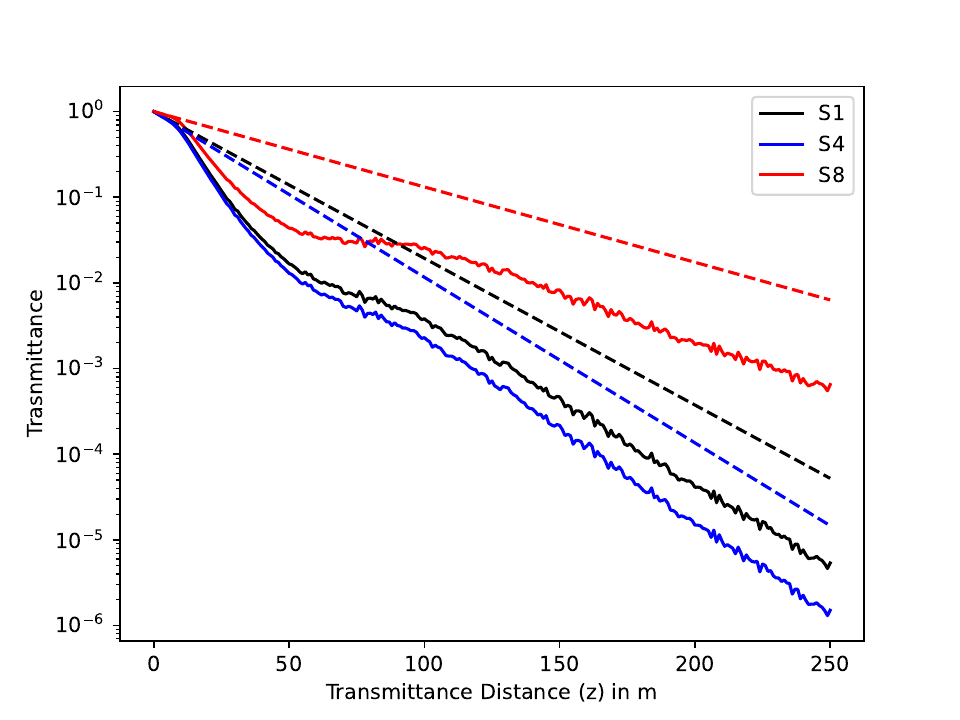}
        \captionsetup{justification=centering} 
        \caption{}
        \label{T_EvET_LOS_S148_1000.png}
    \end{subfigure}
    \hfill
    \begin{subfigure}{0.48\textwidth}
        \centering
        \includegraphics[width=\linewidth]{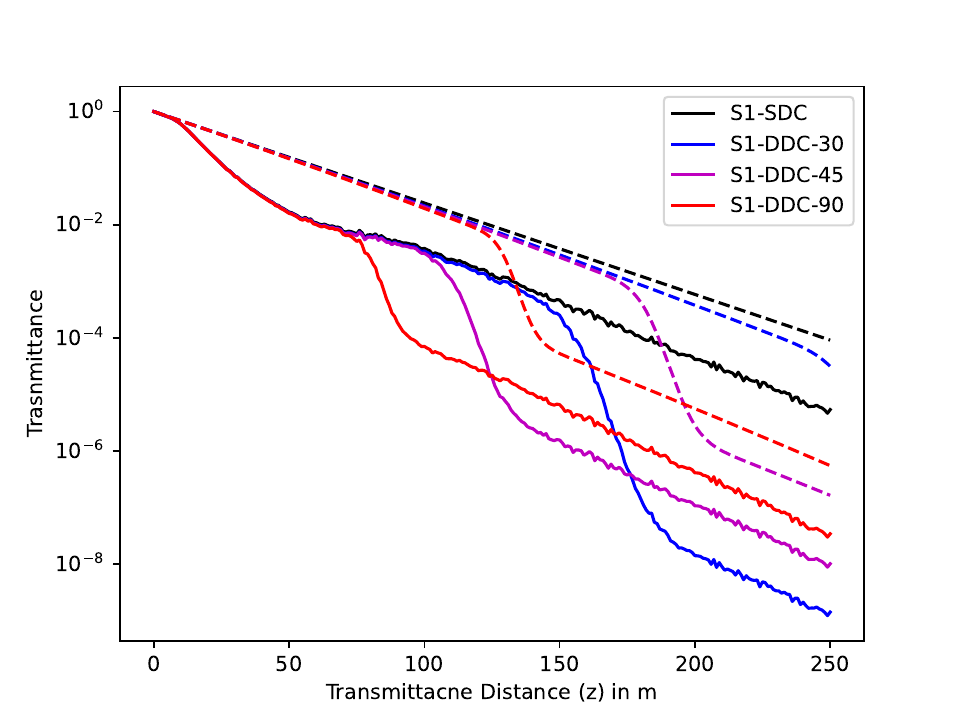}
        \captionsetup{justification=centering} 
        \caption{}
        \label{T_EvET_NOS_1000.png}
    \end{subfigure}
    \hfill
    \begin{subfigure}{0.48\textwidth}
        \centering
        \includegraphics[width=\linewidth]{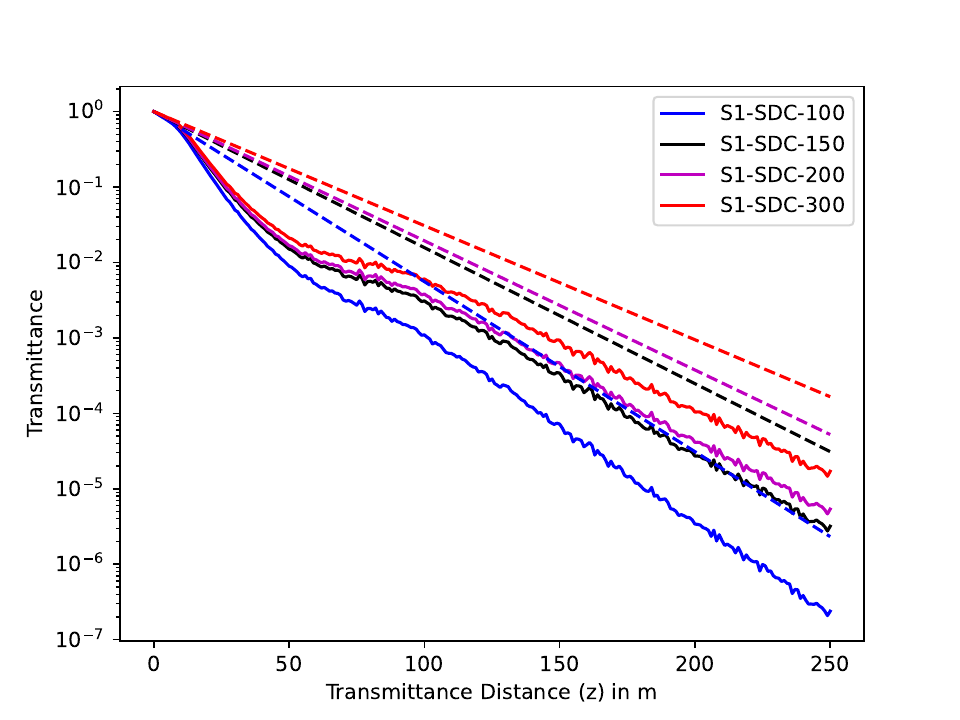}
        \captionsetup{justification=centering} 
        \caption{}
        \label{T_EvET_LOS_S1_depth_1000.png}
    \end{subfigure}

    \caption{(a) Transmittance vs. Transmittance Distance for various sample sizes ($P$) in the presence of turbulence only, (b) Transmittance vs. Transmittance Distance at depth d = 200 meters (SDC) for various types of ocean ($S_1$, $S_4$, and $S_8$), (c) Transmittance vs. Transmittance Distance at various angles (DDC) for type $S_1$ ocean, when transmitter at depth $d=250 m$ and (d) Transmittance vs. Transmittance Distance at various depth levels (SDC) for Type $S_1$ ocean. In  b, c, and d, dashed lines represent effects of attenuation only while solid lines consider the combined effects of attenuation and turbulence.}
    \label{transmittance}
\end{figure*}

In Fig. \ref{transmittance}, we did simulations of the channel transmittance ($T_c$) with respect to the transmittance distance ($z$) between two legitimate parties, exploring various scenarios. In Fig. \ref{transmittance}(a), we specifically simulated the channel transmittance influenced by turbulence alone across different sample sizes ($P=10, 100, 1000$). The figure illustrates that as the sample size increases, the statistical properties of the transmittance data tend to stabilize. Larger sample sizes enable us to encompass a broader range of transmittance values, providing a representation that closely approximates the true transmittance values. For our subsequent calculations, we utilized a sample size of $10^5$ to determine the transmittance. In Fig. \ref{transmittance}(b), we have plotted the transmittance under conditions with and without turbulence. The turbulence conditions are kept the same for all the scenarios. Specifically, for water type $S_8$, which lacks chlorophyll below $d_{\infty}=111.5m$, the attenuation is lower. Comparatively, between water types $S_1$ and $S_4$, $S_4$ exhibits a higher attenuation coefficient value at a depth of $200m$ (refer to Fig. \ref{fig:Attenuation_S1_8}). Consequently, $S_4$ displays lower transmittance than $S_1$. In Fig. \ref{transmittance}(c), we conducted simulations of the transmittance within water type $S_1$ with both legitimate parties positioned at respectively varying depths, corresponding to angles of $0^{o}$, $30^{o}$, $45^{o}$, and $90^{o}$. Notably, Fig.\ref{fig:Attenuation_seawatertype}  reveals a peak region when communication occurs at $30^{o}$, $45^{o}$ and $90^{o}$. At $30^{o}$, where the transmission path is longer (in peak region), the impact is more significant, leading to a notable decrease in transmittance. Similar outcomes are observed for angles of $45^{o}$ and $90^{o}$ degrees, albeit with less pronounced effects. In Figure \ref{transmittance}(d), we simulated the transmittance within water type $S_1$ at various depths ($d=100,150,200,300$). Notably, at 100m depth, there is a higher attenuation coefficient. However, as the depth increases from $150m$ to $300m$ (as depicted in Figure \ref{fig:Attenuation_seawatertype}), there is a corresponding decrease in attenuation, indicating an increase in transmittance with greater depth. In each sub-figure (Fig. \ref{transmittance}), when considering the impact of turbulence on the channel, the transmittance experiences a decline.

In our CV-QKD setup, as depicted in Fig. \ref{fig:QKD_setup}, we implemented the virtual photon subtraction CV QKD protocol. The entanglement source was provided by a TMSV source with a variance of $V = 20$.

\begin{figure*}[ht]  
    \centering
    \begin{subfigure}{0.49\textwidth}
        \centering
        \includegraphics[width=\linewidth]{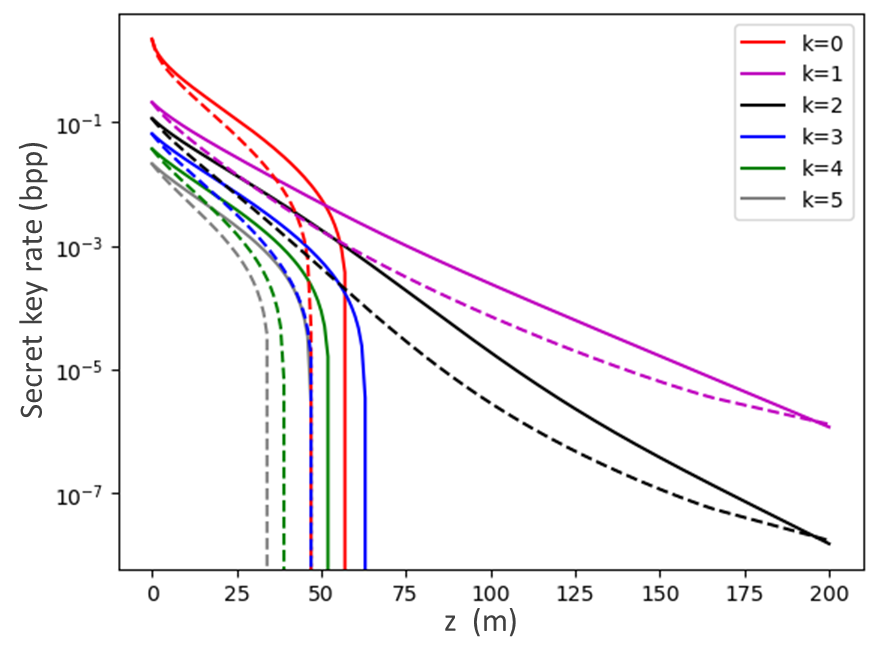}
        \captionsetup{justification=centering} 
        \caption{Keyrate for $S_1$ type water at depth $250m$ without turbulence}
        \label{fig:subplot1}
    \end{subfigure}
    \hfill
    \begin{subfigure}{0.49\textwidth}
        \centering
        \includegraphics[width=\linewidth]{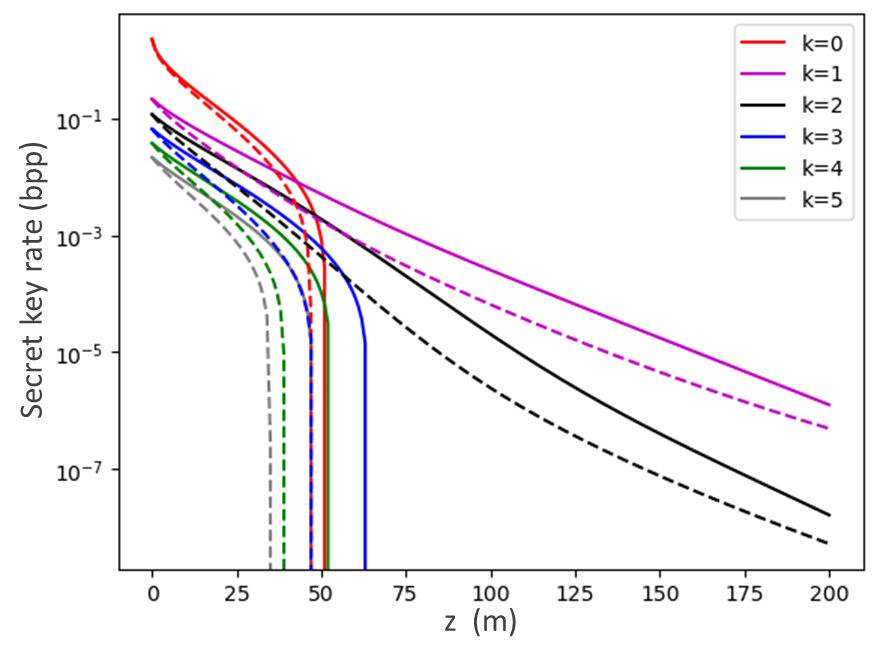}
        \captionsetup{justification=centering} 
        \caption{Keyrate for $S_8$ type water at depth $150m$ without turbulence}
        \label{fig:subplot2}
    \end{subfigure}
    \hfill
    \begin{subfigure}{0.49\textwidth}
        \centering
        \includegraphics[width=\linewidth]{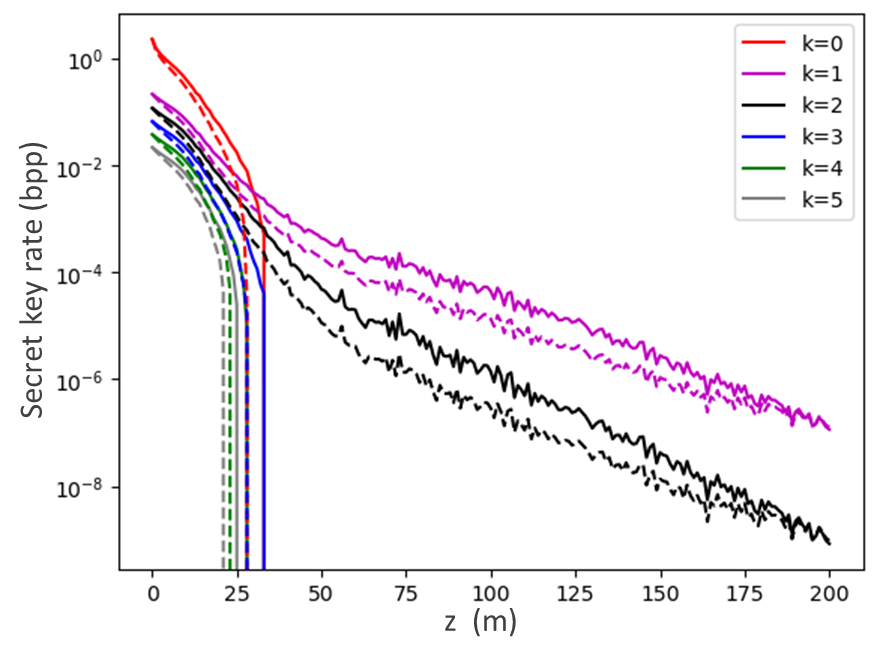}
        \captionsetup{justification=centering} 
        \caption{Keyrate for $S_1$ type water at depth $250m$ with turbulence}
        \label{fig:subplot3}
    \end{subfigure}
    \hfill
    \begin{subfigure}{0.49\textwidth}
        \centering
        \includegraphics[width=\linewidth]{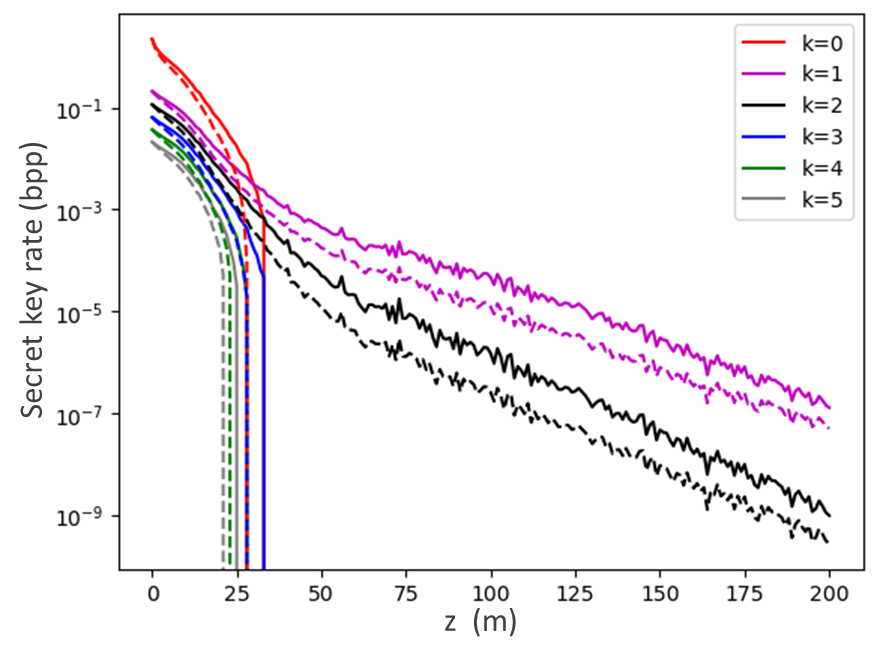}
        \captionsetup{justification=centering} 
        \caption{Keyrate for $S_8$ type water at depth $150m$ with turbulence}
        \label{fig:subplot4}
    \end{subfigure}

    \caption{The maximal secret keyrate as a function of transmission distance ($z$). The red (solid and dashed) lines show the original CV-QKD protocol without photon subtraction. Other lines represent one-photon subtraction (magenta line), two-photon subtraction (black lines), three-photon subtraction (blue lines), four-photon subtraction (green lines), and five-photon subtraction (grey lines) in which solid lines depict SDC and dashed lines for DDC at angle $45^{o}$ (upward). The simulation parameters are as follows: the variance of TMSV state V=20, channel loss parameter ($C_{\text{small}}=0.02$, $C_{\text{large}}=0.01$, $\lambda=480nm$), excess noise $\epsilon=0.01$, and reconciliation efficiency is $\beta=0.95$.}
    \label{fig:overall}
\end{figure*}

For the purposes of our key rate calculations, we considered a reverse reconciliation factor ($\beta$) set at 0.95, an access noise ($\epsilon$) of 0.01, and assumed perfect efficiency of transmitter and receiver detectors ($\eta = 1$). Key rate determination involved the utilization of the channel transmittance equations outlined in  \eqref{channel_transmittance} for different ocean depth scenarios and types of water.
In our analysis, we primarily focused on the comparison of two distinct ocean profiles, namely $S_1$ and $S_8$, in the context of quantum communication key rates. In the case of $S_1$, we assumed the presence of a QKD setup at a depth of $z=200$ meters, while for $S_8$, we considered a depth of $z_1=150$ meters. This comparison was intended to elucidate the influence of varying ocean depths on the key rate.

Additionally, our investigation included a scenario in which both legitimate parties engaged in quantum communication were not situated at the same depth. This particular situation is depicted as a dotted line in the key rate figure. This aspect of our study underscores the impact of the relative depths of the communicating entities within the ocean environment.

In Figure \ref{fig:overall}, we depict a graphical representation illustrating the relationship between the key rate and transmittance distance ($z$) for different scenarios in the context of QKD within our  setup. Each of the subfigures distinguishes between two scenarios: one involving Gaussian QKD (depicted by the red line) and the other involving Non-Gaussian QKD. Notably, for cases where three or less  photon subtractions occur, the key rate surpasses that of the Gaussian approach. For  three or more than three photon subtraction case the tolerable  noise starts to dominate over the correlation feature induced by non-Gaussian operation; as a result the keyrate is unable to surpasses over the Gaussian approach.

Figures \ref{fig:overall}(\subref{fig:subplot1}) \& (\subref{fig:subplot2}) illustrate the influence of attenuation losses alone on the key rate, whereas in Figs. \ref{fig:overall}(\subref{fig:subplot3}) \& (\subref{fig:subplot4}), the effect of turbulence is also taken into account using our proposed model. The decrease in key rate is evidently noticeable as a result of turbulence. In Figs. \ref{fig:overall}(\subref{fig:subplot1}) \& (\subref{fig:subplot3}), we examined the case of water type $S_1$, with Alice situated at a depth of 250m. Conversely, in Figs. \ref{fig:overall}(\subref{fig:subplot2}) \& (\subref{fig:subplot4}), the focus is on water type $S_8$, with Alice located at a depth of 150m. Given that $S_8$ water type contains chlorophyll up to $d_{\infty}=111.5m$, the higher losses in the $S_1$ water type impact the key rate when compared to $S_8$ water type. In each subfigure, both SDC and DDC configurations at $45^{o}$ are considered to observe the effect at different depths (as depicted in the transmittance figure). Notably, the DDC configuration exhibits lower key rates. The channel parameters remain consistent across the subfigures, except for variations in depth and ocean profile.

Within our theoretical and simulation framework, the graphical representation not only highlights the effects of attenuation losses and turbulence loss but also considers the impact of photon subtractions and various communication configurations. These results underscore the considerable potential of our proposed turbulence model, not only for the study of quantum communication protocols but also for optimizing underwater communication channels.

\section{Conclusion} \label{sec:conclusion}

In this work, the primary focus is on optimizing underwater communication channels for quantum key distribution (QKD). The study involves a thorough investigation of various environmental factors, including the selection of wavelength and ocean depth, types of ocean water, the impact of turbulence, and the analysis of channel transmittance.

A comprehensive turbulence model has been constructed, taking into account critical parameters such as Gaussian beam size, transmitter and receiver aperture radii, water salinity, water temperature, and turbulence cell density. The model is visually represented through contour plots and probability density functions, offering insights into the evolving characteristics of turbulence with varying transmittance distances. The channel transmittance simulations explore different scenarios, including the influence of turbulence alone, the impact of turbulence on transmittance, and variations in transmittance with respect to different water profiles, depths, and angles. 

This work highlights the intricate interplay between environmental conditions and quantum communication outcomes. The CV-QKD setup involves the implementation of the virtual photon subtraction CV-QKD protocol. Key rate calculations consider factors such as reverse reconciliation factor, access noise, and detector efficiency. The analysis primarily focuses on comparing key rates for two distinct ocean profiles (S1 and S8) at different depths, shedding light on the influence of varying ocean depths on quantum communication key rates.

Our turbulence modeling provides clarity in signal transmission, thereby enhancing the detection and tracking of underwater threats. In the future, there is potential interest in exploring a quantum communication model that incorporates dynamic underwater conditions, particularly when both the quantum transmitter and receiver are mobile submarines navigating within the ocean.

\section*{Funding and Acknowledgments}

\subsection*{Acknowledgements}
The author would like to thank CSIR for the fellowship support. SB acknowledges support from Interdisciplinary Cyber Physical Systems (ICPS) programme of the Department of Science and Technology (DST), India, Grant No.:DST/ICPS/QuST/Theme-1/2019/6. 

\subsubsection*{Declarations}
The authors declare no conflicts of interest related to this research.

\bigskip

\bibliographystyle{osajnl}
\bibliography{source}

\appendix

\section*{Appendix-A} \label{appendix:a}
\subsubsection*{Absorption}
Different substances in ocean have specific absorption spectra, meaning they absorb light at different wavelengths. For example, chlorophyll, a pigment found in marine plants and phytoplankton, absorbs light primarily in the blue and red parts of the spectrum while reflecting green light, giving the ocean its characteristic blue color. The absorption properties of ocean have significant implications for marine life, as they determine the available light for photosynthesis and affect the distribution of marine organisms in the water column. They are also essential for understanding the optical properties of the ocean, which are critical for applications such as remote sensing, oceanography, and underwater communication. The full absorption spectrum multiplied by there respective concentration, such that \cite{Johnson:13}
\begin{multline}
    a(\lambda, d)=a_w(\lambda)+a_f^{0}(\lambda, d) C_f \exp{\left({-k_f \lambda}\right)} \\
    +a_h^{0}(\lambda, d) C_h \exp{\left({-k_h \lambda}\right)} + a_c^{0}(\lambda,d) \left(C_c/C\right)^{0.602},
\end{multline}
Where:
$a_w$ is the pure water absorption coefficient in $m^{-1}$,
$a_f^{0}$ is the specific absorption coefficient of fulvic acid in $m^{-1}$,
$a_h^{0}$ is the specific absorption coefficient of humic acid in $m^{-1}$,
$a_c^{0}$ is the specific absorption coefficient of chlorophyll in $m^{-1}$,
$C_f$ is the concentration of fulvic acid in $mg/m^3$,
$C_h$ is the concentration of humic acid in $mg/m^3$,
$C_c$ is the concentration of chlorophyll-a in $mg/m^3$,
$k_f$ is the fulvic acid exponential coefficient ($k_f = 0.0189 m^{-1}$), and
$k_h$ is the humic acid exponential coefficient ($k_h = 0.01105 m^{-1}$).

\subsubsection*{Scattering}
The scattering spectra in ocean are impacted by two main biological factors: scattering caused by pure water and scattering resulting from particulate substances. The latter can be further divided into two categories: small and large particles, each exhibiting distinct statistical distributions and scattering properties. These factors collectively contribute to the overall scattering behavior in the ocean, influencing how light interacts with the water and its constituents. The full form equation is given as \cite{Johnson:13}:
\begin{equation}
    b(\lambda, d)=b_w(\lambda)+b_s^{0}(\lambda) C_s(d) +b_l^{0}(\lambda) C_l(d),
\end{equation}
where: The pure water scattering coefficient is represented as $b_w$ in $m^{-1}$,
$b_{s}^{0}$ is the scattering coefficient for small particulate matter in ${m^{2}/{g}}$,
The scattering coefficient for large particulate matter is denoted as ${b_{l}^{0}}$ and is also measured in ${m^2}/g$, ${C_s}$ represents the concentration of small particles in $g/{m^{3}}$, and ${C_l}$ represents the concentration of large particles in $g/{m^{3}}$.

Chlorophyll-a is a predominant substance found in phytoplankton, a group of microscopic organisms. These photosynthesizing organisms thrive in the photic or euphotic zone of the ocean, where sunlight can penetrate. For efficient photosynthesis, phytoplankton require an adequate supply of nutrients, which are usually more abundant in coastal areas due to land runoff and upwelling of subsurface water into the photic zone \cite{raymont2014plankton}. The spectral dependencies for the scattering coefficients of small and large particulate matter are given by the following equations:
\begin{align}
    b_w(\lambda)&=0.005826(400/\lambda)^{4.322},\\
    b_s^0(\lambda)&=1.1513(400/\lambda)^{1.7},\\
    b_l^0(\lambda)&=0.3411(400/\lambda)^{0.3}.
\end{align}
Scattering contributes much less to the overall attenuation coefficient than absorption, through it is much greater in particulate-rich areas.

\begin{figure*}[ht] 
    \centering
    \caption{Parameter Values for S1–S9 Chlorophyll Concentration Profiles (Adapted from \cite{uitz2006vertical}). $C_{chl\text{-}surf}$: Chlorophyll concentration at ocean surface,  $B_0$: Background Chlorophyll concentration on the surface, S: verticle gradient of concentration, h: Chlorophyll above background level, $d_{max}$: depth where Chlorophyll has maximum concentration and $d_{\infty}$: depth where Chlorophyll has negligible concentration.}
    \label{tab:chlorophyll_profiles}
    \begin{tabular}{cccccccc}
        \toprule
        & $\mathbf{C_{chl\text{-}surf}\ (mg/m^3)}$ & $\mathbf{B_0\ (mg/m^3)}$ & $\mathbf{S \times 10^{-3}\ (mg/m^2)}$ & $\mathbf{h\ (mg)}$ & $\mathbf{d_{max}\ (m)}$ & $\mathbf{C_{chl}(d_{max})\ (mg/m^3)}$ & $\mathbf{d_{\infty}\ (m)}$ \\
        \midrule
        S1 & $<0.04$ & 0.0429 & $-0.103$ & 11.87 & 115.4 & 0.708 & 415.5 \\
        S2 & $0.04$-$0.08$ & 0.0805 & $-0.260$ & 13.89 & 92.01 & 1.055 & 309.6 \\
        S3 & $0.08$-$0.12$ & 0.0792 & $-0.280$ & 19.08 & 82.36 & 1.485 & 282.2 \\
        S4 & $0.12$-$0.2$ & 0.143 & $-0.539$ & 15.95 & 65.28 & 1.326 & 264.2 \\
        S5 & $0.2$-$0.3$ & 0.207 & $-1.03$ & 15.35 & 46.61 & 1.557 & 200.7 \\
        S6 & $0.3$-$0.4$ & 0.160 & $-0.705$ & 24.72 & 33.03 & 3.323 & 226.8 \\
        S7 & $0.4$-$0.8$ & 0.329 & $-1.94$ & 25.21 & 24.59 & 3.816 & 169.1 \\
        S8 & $0.8$-$2.2$ & 1.01 & $-9.03$ & 20.31 & 20.38 & 4.556 & 111.5 \\
        S9 & $2.2$-$4$ & 0.555 & 0 & 130.6 & 9.87 & 136.5 & — \\
        \bottomrule
    \end{tabular}
    \label{table:water_profile}
\end{figure*}

Monitoring the near-surface chlorophyll concentration in the ocean is essential, and the NASA SeaWiFS (Sea-viewing Wide Field-of-View Sensor) project employs ocean color observation and quantification to determine chlorophyll concentration \cite{hooker1992seawifs}. Higher chlorophyll concentrations are typically observed along the equator, east-facing coastlines, and in high latitude regions. In the open ocean, typical chlorophyll concentrations range from $0.01$ to $4.0$ mg/m³, while near-shore levels can be as high as $60$ mg/m³.
The chlorophyll profile over a depth $d(m)$ from the surface $C_c(d)$ can be modeled as a Gaussian curve that includes five numerically determined parameters and has the generic form \cite{kameda1998chlorophyll}:

\begin{equation}
    C_c(d)=B_0+\textit{S}d+\frac{h}{\sigma\sqrt{2 \pi}} \exp{\left[-\frac{(d-d_{max})^2}{2 \sigma^2}\right]},
\end{equation}
 where $B_0$ is the background chlorophyll concentration on the surface, $\textit{S}$ is the vertical gradient of concentration, which is always negative due to the slow decrease in chlorophyll concentration with depth, $h$ is the total chlorophyll above the background level and standard deviation of chlorophyll concentration ($\sigma$):
\begin{equation}
    \sigma=\frac{h}{\sqrt{2\pi\left[C_{chl}(d_{max})-B_0-Sd_{max}\right]}}.
\end{equation}

Haltrin proposed a simplified one-parameter model for attenuation, which establishes the relationship between the concentrations of different particulates \cite{pontbriand2008diffuse}. These particulate concentrations were determined numerically in relation to the chlorophyll concentration and are expressed as follows:
\begin{align}
    C_f = 1.74098 \cdot C_c \cdot \exp(-0.12327 \cdot C_c),
\end{align}
\begin{align}
    C_h = 0.19334 \cdot C_c \cdot \exp(-0.12343 \cdot C_c),
\end{align}
\begin{align}
    C_s = 0.01739 \cdot C_c \cdot \exp(-0.11631 \cdot C_c),
\end{align}
\begin{align}
    C_l = 0.76284 \cdot C_c \cdot \exp(-0.03092 \cdot C_c).
\end{align}

Recall that the specific attenuation of chlorophyll also contained a chlorophyll concentration term; so now too has depth dependency :
\begin{equation}
    a_c^0 (\lambda,d)=A(\lambda) C_c(d)^{-B(\lambda)}.
\end{equation}
Where $A(\lambda)$ and $B(\lambda)$ represent empirical constants, a full list of which is given in \cite{bricaud1995variability}.

\section*{appendix-B} \label{appendix:b}
Suppose we have the covariance matrix for the final output state,
\begin{equation}
     V_{out} =
    \begin{bmatrix}
        V_1 I & \phi\sigma_z \\
        \phi\sigma_z & V_2 I
    \end{bmatrix},
\end{equation}
where $I$ is $diag(1,1)$, and $\sigma_z$ is $diag(1,-1)$, and Alice always uses heterodyne detection.
 The mutual information between Alice and Bob can be defined as follows,
 \begin{equation}
     I^{Hom}_{A:B}=\frac{1}{2}\log_2\left(\frac{V_A}{V^{Hom}_{A|B}}\right),
 \end{equation}
where $V_A=(V_1+1)/2$, $V_B=V_2$, and 
\begin{equation}
    {V^{Hom}_{A|B}}=V_A- \frac{\phi^2}{2V_B}=\frac{V_A+1}{2}- \frac{\phi^2}{2V_B}.
\end{equation}
 If we assume that Eve can purify the entire system, the Holevo quantity for homodyne measurements can be expressed as:
 \begin{equation*}
     \chi^{Hom}_{BE}=S(E)-S(E|B)=S(AB)-S(A|B),
 \end{equation*}
where $S(AB)$ is a function of the symplectic eigenvalues $\lambda_{1,2}$ of output matrix , which is 
\begin{equation}
    S(AB) = G\left(\frac{{\lambda_1 - 1}}{2}\right) + G\left(\frac{{\lambda_2 - 1}}{2}\right),
\end{equation}
where 
\begin{equation}
    G(x)=(x+1)\log_2(x+1)-x\log_2{x},
\end{equation}
and 
\begin{equation}
    \lambda_{1,2} = \frac{1}{2} \left(\Delta \pm \sqrt{\Delta - 4D^2}\right).
\end{equation}
Here we have used the notations $\Delta=V^{2}_{1}+V^{2}_{2}-2\phi^{2}$, and $D=V_1 V_2-\phi^2$. Also, $S(A|B)=G\left(\frac{{\lambda_3 - 1}}{2}\right) $ is a function of the symplectic eigenvalue $\lambda_3$ of the covariance matrix $\gamma^{b}_A$ of the A mode after Bob's homodyne detection, where $\lambda_3 = \sqrt{V_1 \left(V_1 - \frac{\phi^2}{V_2}\right)} $. The  covariance matrix $\gamma^{b}_A$ can be described as follow \cite{laudenbach2018continuous}:
\[\gamma^{b}_A=\gamma_A-\frac{1}{V_2}\Sigma_c \Pi_{x,p} \Sigma_c, \]
where $\gamma_A=V_1 I_2$ , $\Sigma_c=V_2  \sigma_z$,  $\Pi_{x}=diag(1,0)$ (in case $x$ is measured) and $\Pi_{p}=diag(0,1)$ (in case $p$ is measured).


\section*{appendix-C: Covariance Matrix Of m-Photon-subtracted TMSV State} \label{appendix:c}
Suppose $\gamma^{(m)}_{AB_2}$ represents the covariance matrix of $\rho^{(m)}_{AB_2}$, and it has the following formula:

\begin{equation}
\gamma^{(m)}_{AB_2} = \begin{bmatrix}
V_A I_2 & \phi_{AB}\sigma_z \\
\phi_{AB}\sigma_z & V_B I_2
\end{bmatrix}
\end{equation}

where, \(V_A\) and \(V_B\) represent the variances in modes \(A\) and \(B_2\), respectively. The covariance between the quadratures of modes \(A\) and \(B_2\) is denoted as \(\phi_{AB}\). The matrices \(I_2\) and \(\sigma_z\) are defined as diagonal matrices with elements \((1,1)\) and \((1,-1)\), respectively.

Suppose $x'$, $p'$ are the heterodyne measurement results of mode A, and $x$ is the homodyne measurement result of mode B. Then,

\begin{align}
    V_A=&\langle\hat{x}^{2}_A\rangle= 2 \int{x'^{2} P(x',p',x) dx' dp' dx}-1, \nonumber\\
    \phi_{AB}=&\langle\hat{x}_A\hat{x}_B\rangle=\sqrt{2} \int{x'x P(x',p',x) dx' dp' dx},\label{eq:avg1}\\
    V_B=&\langle\hat{x}^{2}_B\rangle= 2 \int{x^{2} P(x',p',x) dx' dp' dx} \nonumber.
\end{align}

where
\begin{equation}
    P(x',p',x)=WP({x',p'})|\langle x |\sqrt{T}\alpha\rangle|^{2},
\end{equation}
the weighting function $W$ and probability $P(x',p')$ are described in Section \ref{sec:protocol}.

After simplifying Eq. (\ref{eq:avg1}), the resulting expressions are as follows:

\begin{align*}
    \langle\hat{x}_A^2\rangle &= 2\Bar{V} - 1, \\
    \langle\hat{x}_A\hat{x}_B\rangle &= 2\sqrt{T}\zeta \Bar{V}, \\
    \langle\hat{x}_B^2\rangle &= 2 T \zeta^2 \Bar{V} + 1,
\end{align*}

where $\Bar{V}$ is defined as $\Bar{V}=\int{x'^{2} W\;P({x',p'}) \,dx' \,dp'}$, and further calculations yield:

\[
\Bar{V} = \frac{m + 1}{1 - T\zeta^2}.
\]

Here, \(m\) represents the subtracted photon, \(\zeta\) is the squeezing parameter of the two-mode squeezed vacuum (TMSV) source, and \(T\) is the transmittance of the beam splitter ($BS_1$).

\end{document}